\documentclass[12pt]{article}

\usepackage[T2A]{fontenc}  
\usepackage[utf8]{inputenc}
\usepackage[ukrainian, english]{babel}
\usepackage{graphicx}
\usepackage{xcolor}
\usepackage{microtype}
\usepackage{float}

\begin{document}
\begin{center}
{\bf Machine Learning in Application to Automatic Noise Processing of Solar Spectrograms
}
\end{center}

\begin{center}\bf
I. I. Yakovkin$^{a,*}$, A. O. Bartenev$^{b}$, and N. V. Petrova$^{c}$
\end{center}

\begin{center}
$^{a}$Astronomical Observatory, Taras Shevchenko National University of Kyiv, Kyiv 04053, Ukraine.

$^{b}$Department of Physics, University of Puerto Rico, Mayaguez, Puerto Rico 00680, USA.

$^{c}$Institute of Physics of National Academy of Sciences of Ukraine, Kyiv 03028, Ukraine.
\end{center}

\begin{center} \bf
Abstract
\end{center}

In this paper, a machine learning approach for processing solar spectral data were developed. Its performance was demonstrated using the example of the limb solar flare on July 17th, 1981. The results indicated that machine learning can be effectively utilized to transform between differently digitized spectra, fill gaps in unique experimental data, and undertake spectrum cleaning. Specifically, convolutional neural networks were 
devised to transform between reflective and transmissive scans of a solar flare spectrogram. 
This can be a convenient technique for treating the spectrograms of unique solar events, most notably increasing the proportion of observational spectra that can be further analyzed. Namely, in subsequent research, we will be able to confidently incorporate data such as that captured on the edges of spectrograms, which was previously deemed insufficiently reliable due to limitations of available processing techniques. This will consequently increase the number of spectral lines studied for certain observed events, which is paramount for constructing physical models, as the spectral peculiarities are expected to manifest consistently across different spectral lines.

The developed approach also notably facilitates the detection and removal of impurities in the spectrograms. Previously, each distinct feature in the spectra was manually scrutinized to check its integrity in order to be excluded if identified as an impurity, such as a scratch or a dust particle. By employing the suggested protocol for treating the spectrogram, which includes scanning the spectrogram using multiple distinct techniques and then leveraging machine learning for comparison, the process of excluding impurities can now be automated. Furthermore, the spectrum areas affected by such exclusions can be restored, enabling further analysis. 

\begin{bf}
Keywords: 
\end{bf}
Solar flare spectra, Astronomy image processing, GPU computing, Astronomy data analysis, Spectroscopy

\section{Introduction}
Unique spectroscopic measurements, such as those obtained from limb solar flares or prominences, present significant challenges in terms of data preservation for future analysis. These events are rare and cannot be reproduced, and therefore the measurements obtained from them are valuable and irreplaceable. The spectra of such events can provide insight into the physical processes and conditions in extreme environments, such as the corona of the Sun.

Photographic plates and films remain a valuable tool for capturing and analyzing high-resolution spectroscopic data in various scientific fields \cite{r1,r2,r3}. For example, the Horizontal Solar Telescope at the Astronomical Observatory of Taras Shevchenko National University of Kyiv still uses photoplates to capture the spectrum of solar flares and prominences, due to their large photosensitive area of up to 180 by 240 mm simultaneously \cite{r4,r5}.
It should be noted that the mentioned area of 180 by 240 mm is approximately an order of magnitude larger than that of typical full-frame sensors (36 by 24 mm) and specialized CCD sensors such as E2V-250 (42 by 42 mm) \cite{r6}.

There are several ways to digitalize spectra captured on photoplates and photofilms, including scanned images, digitized photoplates, and digitized photofilms \cite{r7, r8, r9},
which can include complex post-processing pipelines \cite{r10}. The choice of the digitalization method depends on the specific requirements of the research, such as the desired resolution, signal-to-noise ratio, and the size of the area imaged. Generally, scanned images provide the highest resolution and the largest dynamic range, while digitized photoplates and photofilms provide a convenient and low-cost alternative. While some digitization methods can be better-fit than other, changing the digitalization method can often be technically challenging, especially when working with such complex systems as the Echelle spectrograph of the Horizontal Solar Telescope. During digitization process, the noise and/or artifacts can occur, which can be dealt with by introducing specific models for such noises \cite{r11, r13}. For instance, in \cite{r12} the portions of image related to the reflected and transmitted rays were separated using the measurements in multiple polarizations and applying an encoder-decoder-type neural network.

Spectroscopic measurements, including those captured on photoplates, can now be processed with improved accuracy thanks to advancements in technology and the integration of machine learning 
(ML) \cite{r14,r15,r16,r17}. Machine learning is a branch of artificial intelligence concerned with the design and development of algorithms that can learn from and make predictions on data. Machine learning has emerged as a powerful tool in the field of computer science, and its applications in image processing and spectroscopy have been rapidly growing \cite{r18, r19, r20}.
ML algorithms can use statistical models to analyze and identify patterns in large amounts of data, which allows them to make predictions and generalizations based on this information with a high level of accuracy and speed. Moreover, modern research has also proposed the methods for applying machine learning to limited amounts of data \cite{r21}. 

In spectroscopy, the application of ML algorithms has been relatively recent, but is already showing great promise \cite{r22,r23,r24,r25}. In particular, ML has found success in improving image quality and remove noise, as well as to enhance images by increasing their resolution or removing artifacts \cite{r26,r27}. Convolutional neural networks (CNNs) have proven particularly effective for these tasks, and have been applied to a wide range of image data, from natural images to medical imaging \cite{r28,r29}. 
These capabilities make ML a valuable tool in spectroscopy, specifically with regards to its ability to transform between differently digitalized spectra, and fill the gaps in unique experimental data \cite{r30,r31,r32}. 
This study employs machine learning to transform reflective scans into transmissive scans and vice versa, aiming to identify whether reflective scans can provide additional insights for processing transmissive scans and detecting spectral contamination.

\section{Material}
The main goal of this research was to enhance the accuracy and precision of processing unique spectral data, using the example of the limb solar flare that occurred on July 17th, 1981. The observed spectra were captured on the photoplates with the Echelle spectrograph of the Horizontal Solar Telescope by N.I. Lozitska, V.G. Lozitsky, and P.M. Polupan \cite{r5} and present important insights into the nature of the magnetic fields associated with solar flares \cite{r4,r5,r33}. The challenge with these spectra is that they are one-of-a-kind measurements and can only be recorded once, therefore it is crucial to obtain the most accurate results from processing them.

\begin{figure}
\begin{center}
\includegraphics[width=100mm]{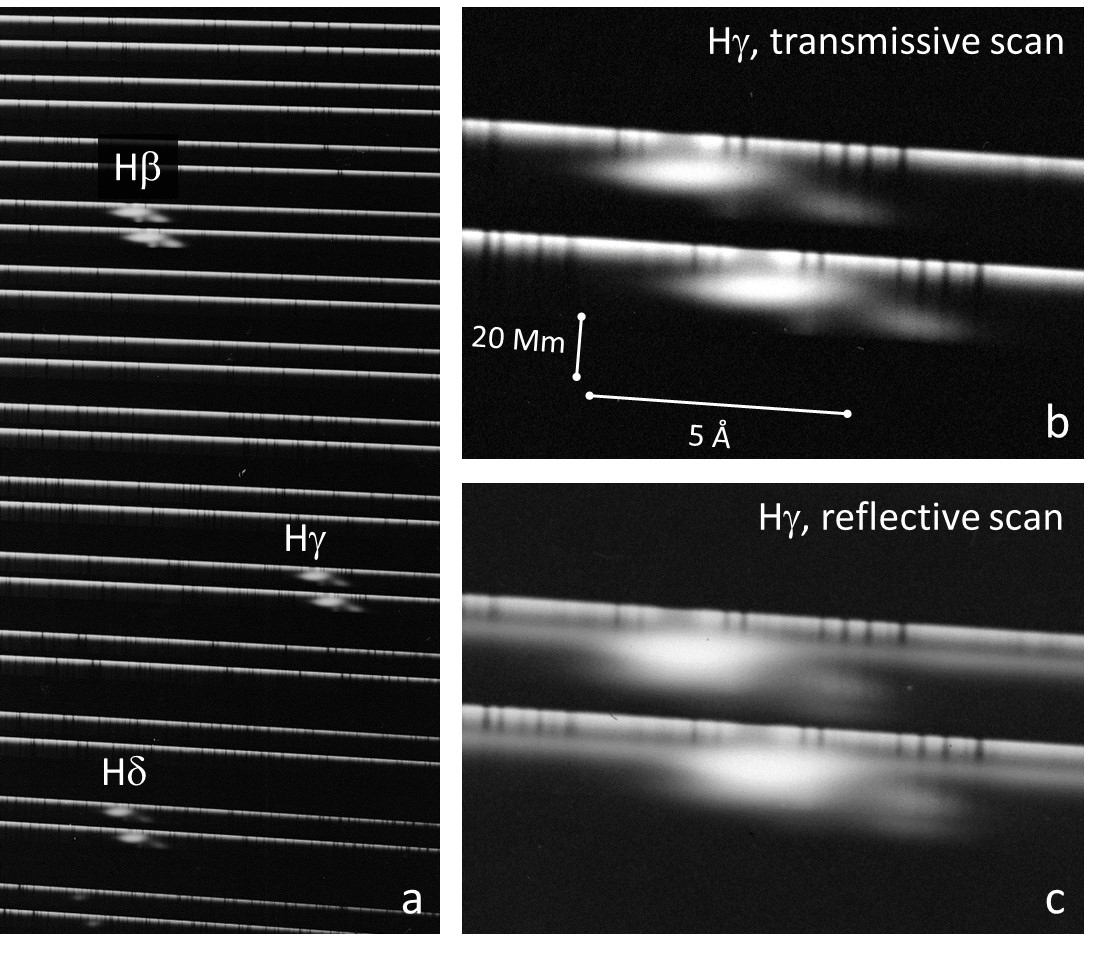}
\caption{Spectral images used as inputs for the neural networks. A portion of a photoplate featuring bright emissions in three spectral lines (H$\beta$, H$\gamma$, H$\delta$), represented by pairs of bright spots corresponding to opposite circular polarizations (a). The input data for the study included high-resolution reflective (c) and transmissive (b) scans of eight spectral lines. The wavelengths increase towards the right, while the height above the photosphere increases downward. While transmissive scans are known for their accuracy, reflective scans may contain additional reflections that cause a blurred and shifted "halo" in the spectrogram. }
\label{f1}	
\end{center}
\end{figure}

Figure \ref{f1}a shows a portion of the 8:33 UT photoplate where bright emissions in 3 spectral lines (H$\beta$, H$\gamma$, H$\delta$) can be seen. 
The Figure captures the spectrum in 13 diffraction orders, each order being represented by a pair of horizontal stripes corresponding to two circular polarizations (the cross-dispersion mechanism of the Echelle spectrograph is discussed in more details in \cite{universe}).
Within each stripe, the wavelength increases towards the right, and the height above the photosphere increases downward. It should once again be emphasized that different stripes correspond to different diffraction orders of the spectrograph, meaning that there is no continuous height or wavelength scale across the entire photoplate.

Figure \ref{f1}b presents an enlarged portion of Figure \ref{f1}a, showing the details of the spectrum near the H$\gamma$ line in two circular polarizations obtained using the transmissive scanning technique. The local height and wavelength scale references are marked by the vertical and horizontal intervals, respectively. Figure \ref{f1}b shows the same portion of the spectrogram as Figure \ref{f1}b, but obtained using a reflective scanning.
The entrance slit of the spectrograph was positioned in such a way that it captured a part of the solar limb (for more details see \cite{r4}). Consequently, the upper portions of each stripe correspond to the continuous spectrum of the photosphere with Fraunhofer absorption lines. On the other hand, the lower portions of each stripe correspond to the chromosphere and lower corona, where emission in only a few selected lines is present.


The training dataset consisted of high-resolution reflective and transmissive scans of 8 spectral lines. These images had the same resolutions, but were of significantly different sizes, with the average dimensions being approximately 1000 by 1000 pixels. All images are grayscale and are represented as 2D arrays with values between 0 (black) and 1 (white). The data corresponds to scans of 4 photoplates that capture 8:17 UT, 8:33 UT, 9:02 UT and 9:51 UT moments of time, respectively. The photoplates are of the same type -- WP1.
The test set consisted of the H$\gamma$ line at 8:33 UT and H$\beta$ line at 9:51 UT. The exposures were different for each moment, ranging within 10 -- 30 seconds. However, the exposure times were selected specifically to account for the different brightness of the flare as it evolves, to ensure that the blackening of the photoemulsion is within the most reliable range of its characteristic curve. Therefore, in the resulting scans of the spectrograms, the brightness ranges are similar for all used material.
Transmissive scans (Fig. \ref{f1}b) have been extensively verified to have good accuracy compared to traditional photometers. 
On the other hand, reflective scans (Fig. \ref{f1}c) of the same photoplate often contain defects such as additional reflections, which result in a blurred and shifted "halo" in the spectrogram. Nevertheless, in this study we show that reflective scans can still provide additional insights for processing transmissive scans.

\section{Neural network design}

To tackle these challenges, Convolutional Neural Networks (CNNs)
were used to perform the transformation of spectra between reflective and transmissive scans. The CNN was implemented in Python using the PyTorch deep learning framework, with GPU acceleration provided by CUDA. During the training process, the CNNs were instructed to learn the relationship between the two types of scans and perform the transformations in both directions (“reflective-to-transmissive” and “transmissive-to-reflective” models). CNNs were chosen to be used for
this task as it has proven to be effective in image processing tasks and can learn to capture the underlying patterns in the data with a relatively low probability of overfitting \cite{lin2013network}.


\begin{figure}
\begin{center}
\includegraphics[width=120mm]{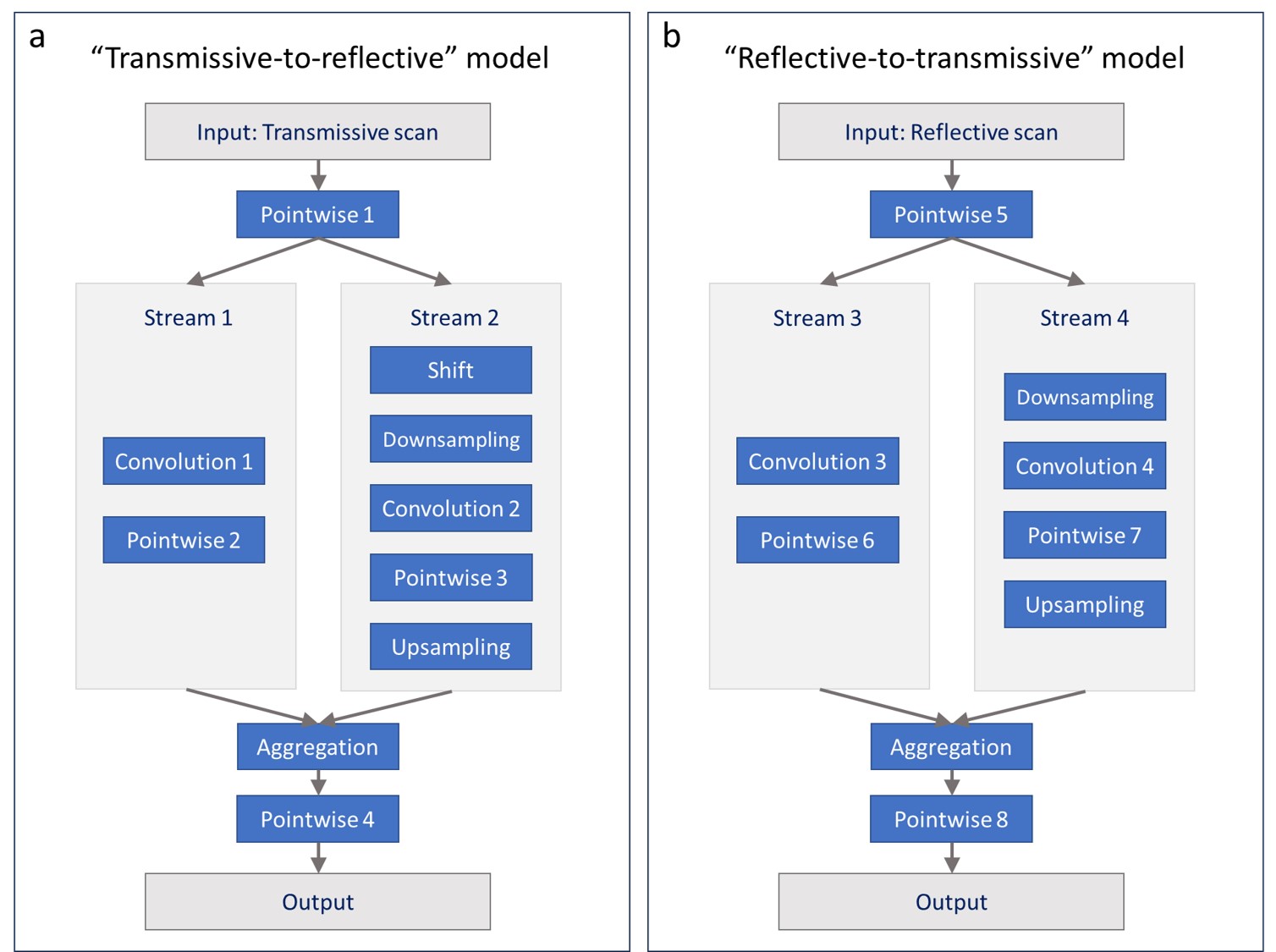}
\caption{Neural Network Architectures for converting between reflective and transmissive spectra scans. Architecture of the "transmissive-to-reflective" model (a) and architecture of the "reflective-to-transmissive" model (b). To optimize the performance of the neural network, 
each model was constructed using two separate streams: one for reproducing the spectrum itself, and the other for adding or removing the halo. By explicitly adding the shift block, the kernel sizes could be decreased and performance in the "transmissive-to-reflective" model improved.}
\label{f2}	
\end{center}
\end{figure}

When designing the neural networks
(Fig. \ref{f2}), various aspects were considered to ensure optimal performance. After visually examining the input and target data, it was determined that it would be beneficial to introduce two separate streams in each model. This decision was based on the observation that the reflective scan is essentially a combination of the transmissive scan and a smoothed version of the same transmissive scan that has been shifted. Since the “transmissive-to-reflective” model (Fig. \ref{f2}a) is much more straightforward (adding a “halo” is easier than removing it), the shift block was explicitly included to one of the streams, which allowed for a significant decrease in kernel sizes, resulting in much better performance. In contrast, “reflective-to-transmissive” model (Fig. \ref{f2}b) is more complicated, and the explicit addition of the shift block showed almost no effect, therefore the choice to stick with large convolution kernels was made.

All convolutions were used with the padding mode set to repeat the nearest values, and the padding sizes were selected for each kernel in the way that retained the input image size. 
The kernel sizes were 5 by 5 for convolutions 1 and 3, 30 by 30 for convolution 2, and 60 by 60 for convolution 4. Each convolution had 15 channels, and no stride or dilation were used.

The shift block is a 2-parameter transformation that shifts the image horizontally and vertically by $s_x$ and $s_y$ pixels, respectively. The shifts can be non-integer, in which case the bilinear interpolation is used to account for sub-pixel shifts. 
The implementation of the shift block and its gradient backward propagation function were created from scratch, however it should be noted that a partial affine transformation could have been used instead without noticeable difference.

Multiple pointwise operations blocks were added to the models, consisting of nonlinearities alternated with linear transformations, to allow for learning a significantly nonlinear intensity relation between differently scanned spectra.
Specifically, the pointwise blocks were implemented as chained sigmoids and linear transformations:
\begin{equation}
 \bigcirc_{i=1}^{5} \lambda x. \sigma(a_i x + b_i),
\end{equation}
where $\bigcirc$ denotes the functional composition operator, $\lambda$ is the lambda abstraction operator, $\sigma$ is the sigmoid function, $a_i$ and $b_i$ are learnable model parameters, $i=\overline{1..5}$. These pointwise blocks therefore play the role of tunable activation layers.

Since the “halo” in the reflective scans is smooth and lacks details, the Convolutions 2 and 4 (Fig. \ref{f2}) were wrapped in a 2x downsampling-upsampling pair, which drastically improved the computation time. For aggregation, the values of the stream outputs were simply added together. It should be noted that we do not assume any specific model for the halo, but rather allow the models to capture the relevant patterns.

The pixelwise loss has been defined as the square of the pixel-wise differences between the output and target images:
\begin{equation}
    L_{i,j} = (o_{i,j}-t_{i,j})^2,
\end{equation}
where $L_{i,j}$ is the pixelwise loss at the location $(i,j)$, $o_{i,j}$ and $t_{i,j}$ are the output and target intensities at the same location. The pixelwise loss, similar to the input, output, and target, has the values between 0 and 1.
The total loss is defined as the Root Mean Square loss with a correction for possible outliers:
\begin{equation}
    L = \sqrt{\langle \textrm{bottom 99.9\% of } L_{i,j} \rangle}.
\end{equation}
In this expression, 0.1\% of the pixels with the highest pixelwise loss values are not taken into account. This correction was made to allow the models to focus on the uncontaminated parts of the images, which are represented by the absolute majority of pixels, and ignore the outliers that are caused by spectral contamination such as scratches and dust particles. 

In terms of the optimization algorithm, various options were considered, such as Stochastic Gradient Descent (SGD), Adaptive Moment Estimation (Adam), and Resilient Propagation (RProp). Through a series of tests and evaluations, it was found that alternation between the RProp and Adam algorithms every 10000 epochs worked best \cite{r34, r35}.
The optimal values of the neural network parameters were obtained through extensive experiments and used to design the final network architecture.
It should be noted that the design of the models significantly limits the likelihood of overfitting due to the usage of CNNs with kernels much smaller than the image sizes. Keeping this in mind, we changed the procedure from the typical \emph{train$\to$validate$\to$test} in favor of \emph{optimize model$\to$train$\to$test}, which is similar to the progressive validation approach \cite{r36}. The first stage was model architecture optimization. During this stage, different hyperparameters of the model (such as kernel sizes, dilations, shift layer, downsamping, etc.) were varied and assessed based on relatively short model training sessions. Next, the model was trained using a single extensive run of 200000 epochs. Finally, the performance of the model was assessed using the test set.

\section{Results}
\begin{figure}
\begin{center}
\includegraphics[width=120mm]{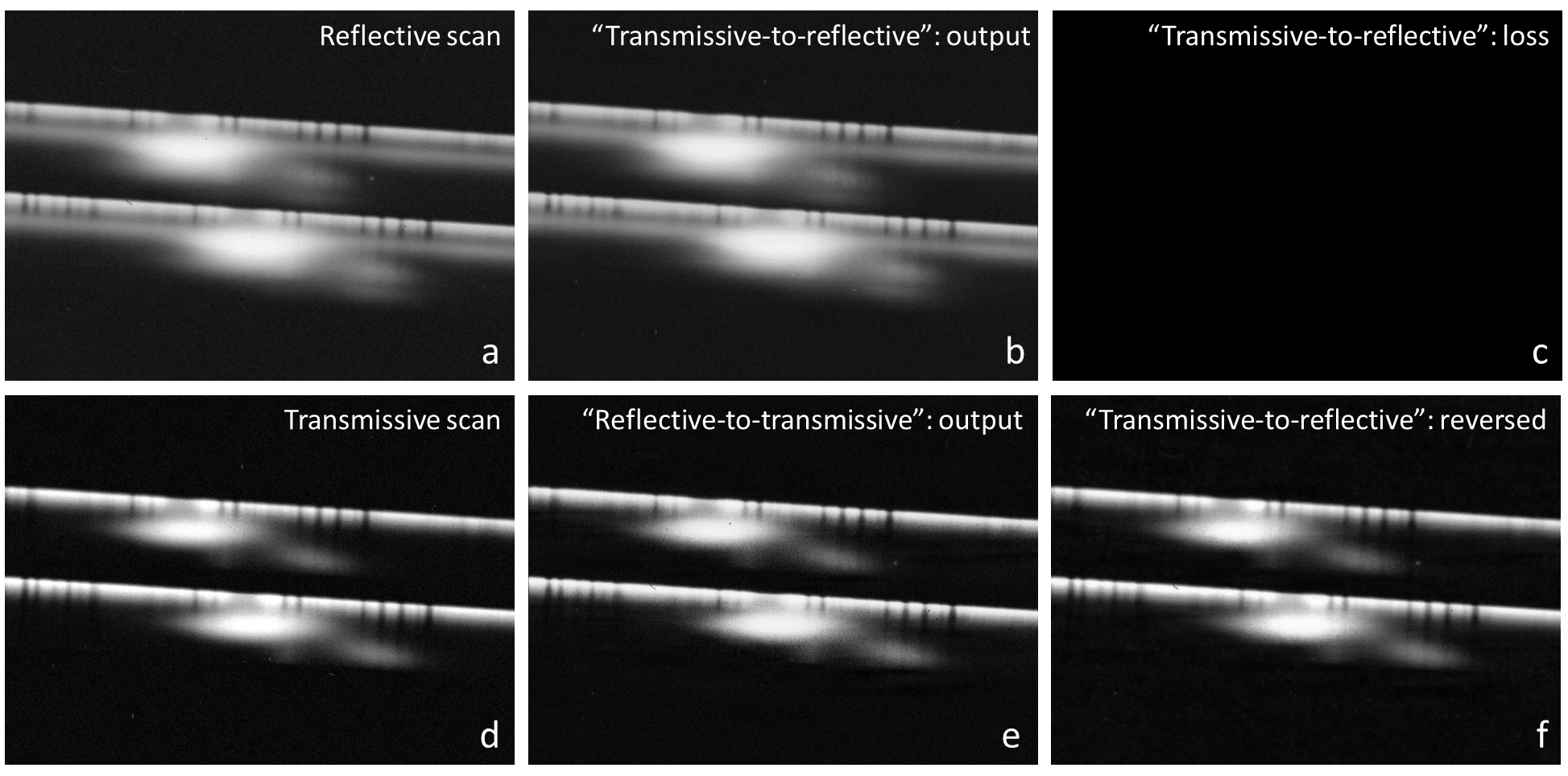}
\caption{Results of training: “transmissive-to-reflective” model target (reflective scan) (a), output (b), the corresponding pixelwise loss $L_{i,j}$ (c), “reflective-to-transmissive” model target (d), “reflective-to-transmissive” model output (e), and output of the reversed “transmissive-to-reflective” model (d).}
\label{f3}	
\end{center}
\end{figure}

When comparing the results of the transformation with the target data, it was evident that the CNN was able to match the target data with high accuracy, both in terms of intensity and positional information of the spectral features. Figure \ref{f3} presents the performance of the "transmissive-to-reflective" and "reflective-to-transmissive" models on the H$\gamma$ line at 8:33 UT, which belongs to the testing set.
The Figure includes the target scans (a and d), model outputs (b and e), and the output of the reversed "transmissive-to-reflective" model (f). The pitch-black pixelwise loss $L_{i,j}$ (c) indicates a good match between the output and the target, with typical total loss $L$ being around 1\%  for the “reflective-to-transmissive” model and less than 0.5\% for the “transmissive-to-reflective” model. This demonstrates the ability of the model to accurately convert the spectra scans from transmissive to reflective and vice versa.

When converting from reflective to transmissive scans, there are two different approaches to consider: learning to transform directly from reflective to transmissive scans (Figure \ref{f3}e) or first training a “transmissive-to-reflective” model and then reversing it (Figure \ref{f3}f). The study has revealed that the latter approach, involving first learning to convert from transmissive to reflective scans and then reversing the network, results in significantly lower computation times
compared to directly learning to transform from reflective to transmissive scans, while maintaining a similar overall accuracy.

\begin{figure}
\begin{center}
\includegraphics[width=100mm]{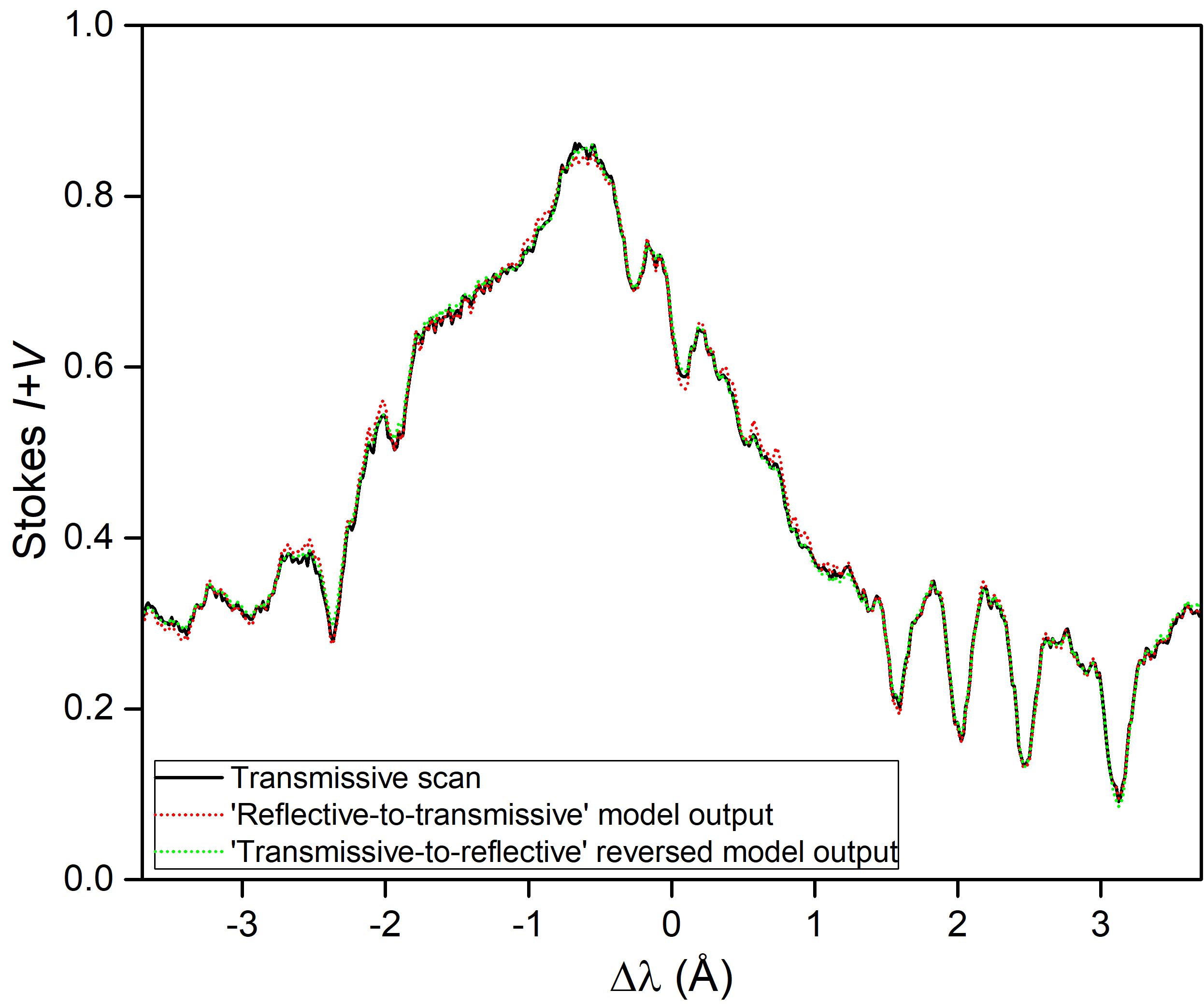}
\caption{Wavelength dependencies of the Stokes $I+V$ parameter obtained from the data on Figure \ref{f3}d, e and f at the height of 0 -- 2 Mm above the photosphere. Despite different digitalization techniques used, the resulting profiles match well.}
\label{f4}	
\end{center}
\end{figure}

Figure \ref{f4} presents the comparison of the Stokes $I+V$ profiles obtained at the heights of 0 -- 2 Mm above the photosphere from the model outputs presented in Figure \ref{f3}e, and f, and the target data (Figure \ref{f3}d). The Stokes $I+V$ profiles are the intensity profiles corresponding to the upper stripes within the spectral images. 
Both the "reflective-to-transmissive" and the reversed "transmissive-to-reflective" models exhibit an excellent agreement with the target transmissive profile, preserving all its distinct features.

\begin{figure}
\begin{center}
\includegraphics[width=120mm]{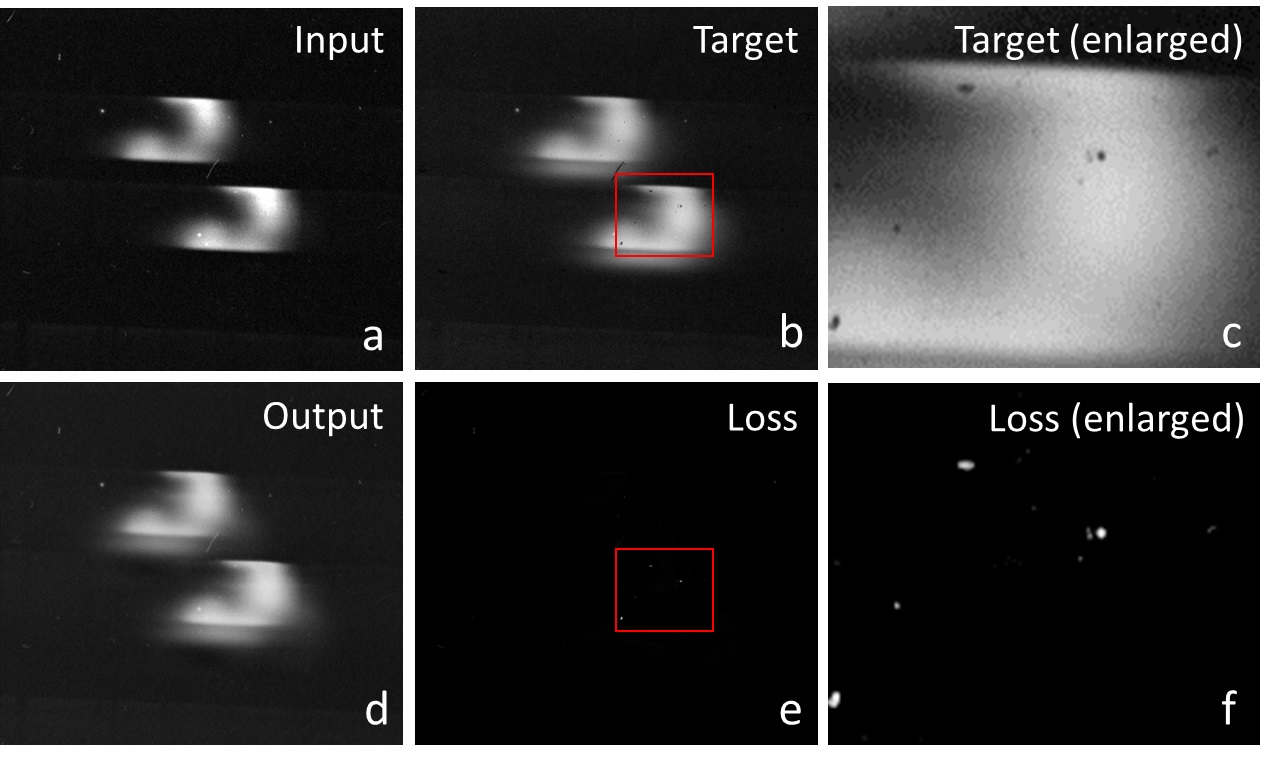}
\caption{Visual representation of the models' performance on spectra with impurities or contamination. The figure shows model's input (a), target (b), its enlarged fragment (c), model output in panel (d), the intensity-normalized pixelwise loss $L_{i,j}/\max(L_{i,j})$ in panel (e), and an enlarged view of the intensity-normalized pixelwise loss of the same area (f). By comparing the panels (c) and (f), it becomes apparent that the areas of high loss correspond to the impurities on the spectral images. This allows to reliably detect and exclude such impurities from the spectra during post-processing.}
\label{f5}	
\end{center}
\end{figure}

Figure \ref{f5} provides a visual representation of the CNN models' performance when dealing with contaminated spectra of the H$\beta$ line at 9:51 UT, belonging to the testing set.
The figure shows the input to the model (a), the target (b), and it’s enlarged fragment (c). The output (d) matches 
the target well, which is evident from the intensity-normalized loss (e). To inspect the loss more closely, an enlarged view of the same area is presented in Figure \ref{f5}f. By examining Fig. \ref{f5}c and f, it is possible to observe that the areas where the loss indicate the presence of impurities.

Enlarged and 10 times intensity-amplified image of the pixelwise loss is presented in Appendix A, Figure \ref{fa1}. Under such amplification, the dust particles and other impurities manifest as bright spots, while the barely noticeable smooth noise variations in the background correspond to the imperfections of the transformation from the transmissive to the reflective scans. Figure \ref{fa2} features the same pixelwise loss, but with even stronger 40-times amplification. The smooth contours of the spectra now become apparent, suggesting that the model had the worst performance near the areas of fast intensity variation.
Additionally, an interesting phenomenon emerges: a long particle, marked with the yellow arrows, appears twice in the pixelwise loss. This is due to the fact that this particle has shifted between the moments when the transmissive and reflective scans were taken. This particle therefore appears in different locations in the transmissive and reflective scans, resulting in two imprints in the pixelwise loss. The imprint on the right, which relates to the transmissive scan, has a low pixelwise loss magnitude, likely due to the impact of the high-intensity halo in the same location.   
By introducing a threshold for the pixelwise loss magnitude, such as $L_{i,j}/\max(L_{i,j})=0.025$, the spectral contaminations become clearly separated (Appendix A, Figure \ref{fa3}). This allows to automatically determine the contaminated locations.

In order to assess the quality of automatic noise detection, a comparison with manual detection was performed. 
At present, the primary method used to remove the spectral contamination from the solar spectrograms involves a careful manual inspection of both the scans of the spectrograms and the spectrograms themselves, and marking the locations of the features that should be removed. This procedure was performed with the reflective and transmissive scans of the H$\beta$ line scans at 9:51 UT (Appendix A, Figures \ref{fa4} and \ref{fa5}, respectively), where the manually detected locations of contamination were marked by red and blue rectangles, respectively.
Next, the locations of the manually detected impurities were compared to the locations of the above-threshold pixelwise loss (Appendix A, Figure \ref{fa6}) by overlaying Figures \ref{fa3}, \ref{fa4}, and \ref{fa5}.
In Figure \ref{fa6}, the locations of particles detected manually in reflective and transmissive scans are marked with red and blue solid rectangles, respectively. The rectangles within which no above-threshold pixelwise loss is present are marked with yellow crosses.
Several locations had above-threshold pixelwise loss, but did not correspond to the locations where the particles were initially manually identified. Such locations were then once again carefully inspected in the original scans, and the presence of contamination was confirmed in either reflective or transmissive scan, as marked with dashed red and blue rectangles, respectively.

Overall, out of 50 distinct locations of above-threshold loss, 30 locations were detected manually. The other 20 locations were not detected initially, but each one was confirmed to be correct post-factum. Out of 38 distinct manually detected features, 8 were not present in Figure \ref{fa3}, as marked by yellow crosses in Figure \ref{fa6}. Nevertheless, these features are present in Figure \ref{fa2}, although with an insufficient intensity to pass the threshold. Therefore, reducing the threshold value may lead to even better performance, but fine-tuning the automatic noise detection is outside the scope of this study.

\section{Discussion}

The models have shown the ability to convert between the reflective and transmissive scans of solar spectrograms. While the dataset contained a limited number of image pairs (8 for training and 2 for testing), it should be noted that each image contains a significant number of independent areas. Indeed, the pixels that are far enough from a selected pixel do not contribute to the output due to the limited kernel sizes. For example, the "transmissive-to-reflective" model, given the Convolution 2 kernel size of 30 pixels, downsampling of 2 times, and vertical shift of approximately 40 pixels, has a sensitivity window of approximately 60 by 100 pixels. This means that a single 900 by 1000 pixels image can be cropped into 150 completely independent samples, and this number can be increased even further by allowing for partial overlap, since the contribution of pixels diminishes near the edges of the sensitivity window. However, due to the nature of convolutions, such explicit cropping is unnecessary, especially considering the increased effect of padding near the edges of the images.

Several notes should be made about the \emph{optimize model$\to$train$\to$test} procedure used. Such an approach is rarely used because it is often deemed unfeasible due to the typically high likelihood of a poor model performance on the test set due to overfitting. However, it was observed that the proposed models are not prone to overfitting even when learning on a single image. This allowed us to significantly expedite the model architecture optimization step, and also to use a higher percentage of available data for actual learning. In the end, the test set showed acceptable performance with no signs of overfitting, therefore completely justifying the chosen approach in this specific case.

During spectral analysis, the data is typically sliced into spectral line profiles corresponding to different positions in space. This approach enables the utilization of a wide array of tools for profile processing, facilitating accurate measurements of physical quantities such as magnetic fields, temperatures, turbulent velocities, and so forth. Figure \ref{f4} demonstrates that the Convolutional Neural Networks accurately maintain all main characteristics of the spectral profile, including positions, heights, widths, and specific shape features of both the emissive and absorptive peaks.

When the models are applied to spectra containing impurities or contamination (Figure \ref{f5}), interesting results emerge. The models appear proficient in reproducing pixels that represent the true darkness of the photoemulsion, which make up the majority of the pixels. 
This leads to very low loss for pixels representing the correct darkness of the photoemulsion, but very high loss for scratches and impurities.
This approach can be used for automatic identification of the scratches, dust particles, and other impurities from the final spectrogram scans, ensuring that the data used for analysis is free from interference. Additionally, the use of reflective scans allows for the detection of impurities that may be hard to identify using transmissive scans alone, as the difference in the relationship between transmissive and reflective scans causes impurities to stand out more prominently (Appendix A). Introducing a threshold for the pixelwise loss allows to conveniently separate all contamination. While assessments done using a single image are insufficient for definitive statistical conclusions, the highlighted approach for automatic noise detection is clearly promising.

\begin{figure}
\begin{center}
\includegraphics[width=100mm]{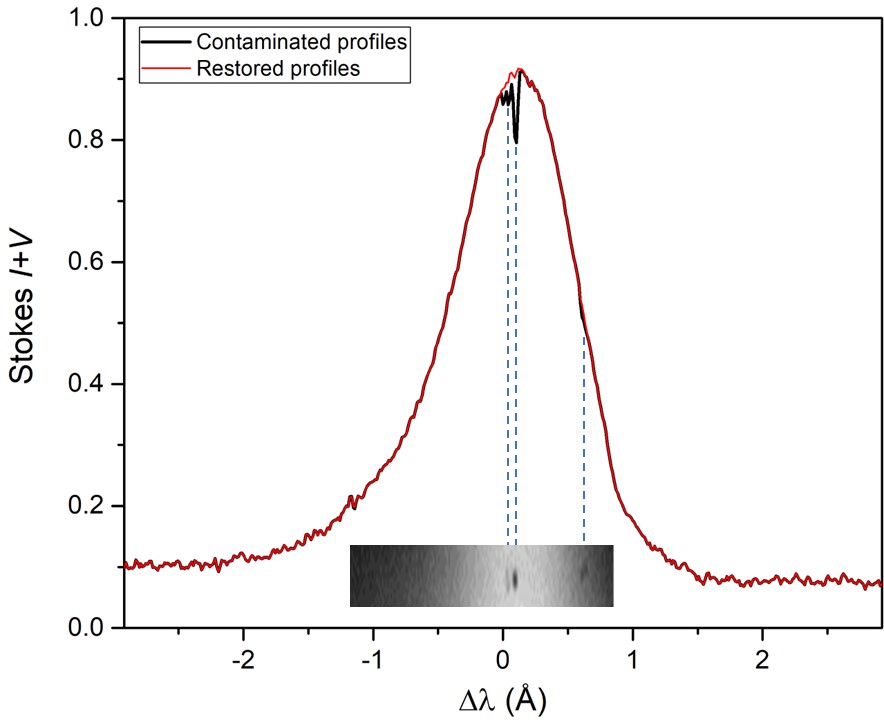}
\caption{Restoration of spectral line profiles with impurities. The inset shows the corresponding portion of the spectral image. The exclusion of contaminated pixels, as detected by the Convolutional Neural Network, allows for the reliable restoration of spectral line profiles. This process preserves the maximum amount of original data and avoids the introduction of unnecessary assumptions.}
\label{f6}	
\end{center}
\end{figure}

Next, we will introduce a protocol for restoring contaminated spectral intensity profiles. Obtaining intensity profiles from spectral images involves a two-step process. The spectral images are first carefully divided into thin strips, each representing a distinct photometric section. This division allows scientists to methodically study different sections of the spectral image in a detailed manner. Once divided, the next step involves deriving intensity profiles by averaging values across the height of each strip. During this process, contaminated pixels, identified using Convolutional Neural Networks,
are excluded from the averaging process (Figure \ref{f6}). 

This technique provides a robust and reliable approach to restoring profile shapes. It avoids unnecessary assumptions and data loss typically associated with simpler methods, such as invalidating entire profile segments and then interpolating the data. The proposed procedure for removing spectral contamination
focuses on making the best use of the original data, and as such, delivers more accurate and trustworthy results.

\section{Conclusions}
This study presents a machine learning-based approach that introduces a 
new technique for processing solar spectrograms. Central to our method is the transformation between solar spectra that have been digitized 
using different scanning techniques. This transformation increases the proportion of analyzable data within captured spectra, enabling, in certain instances, the processing of a greater number of spectral lines in the spectrogram.

Furthermore, we introduce a protocol for automatic impurity detection and removal. By performing an automatic comparison of the differently digitized spectra, impurities emerge as areas with high loss. These areas can then be excluded from the images, and the spectral line profiles restored through weighted averaging.

The practicality of our approach is underscored by its successful application to the spectrograms of the solar flare on July 17th, 1981, where notable impurities were efficiently identified and eliminated. While our method is specifically designed for solar spectra, its potential extends to broader applications in general spectroscopy.

\section*{Acknowledgements}
The authors would like to express their gratitude to V.G. Lozitsky for generously providing the spectra used in this study. 
The research was partially supported by the Taras Shevchenko National University of Kyiv, project No. 22БФ023-03.

\clearpage
\section*{Appendix A}
\renewcommand\thefigure{A\arabic{figure}}   
\setcounter{figure}{0}

\begin{figure}[H]
\begin{center}
\includegraphics[width=120mm]{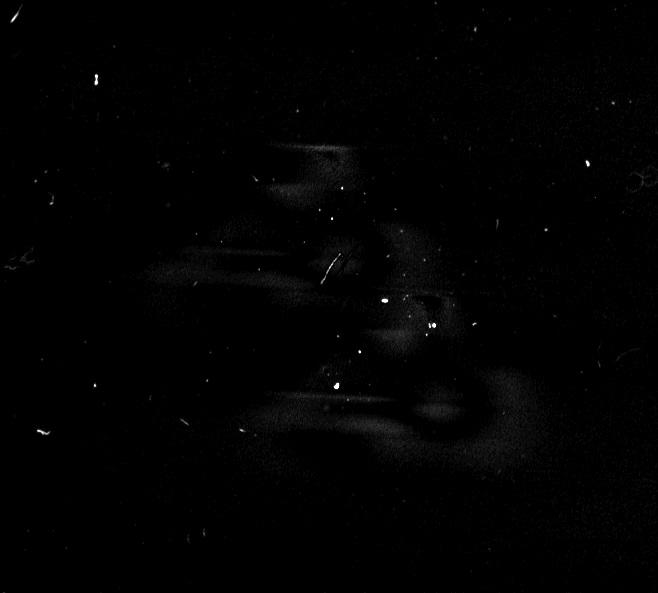}
\caption{Same as in Figure \ref{f5}e, but the brightness in this Figure corresponds to $10 L_{i,j}/\max(L_{i,j})$, capped at 1. At such pixelwise loss amplification, the dust particles and scratches are clearly visible, and smooth contours of the spectral emission are noticeable in the background.}
\label{fa1}	
\end{center}
\end{figure}

\begin{figure}[H]
\begin{center}
\includegraphics[width=120mm]{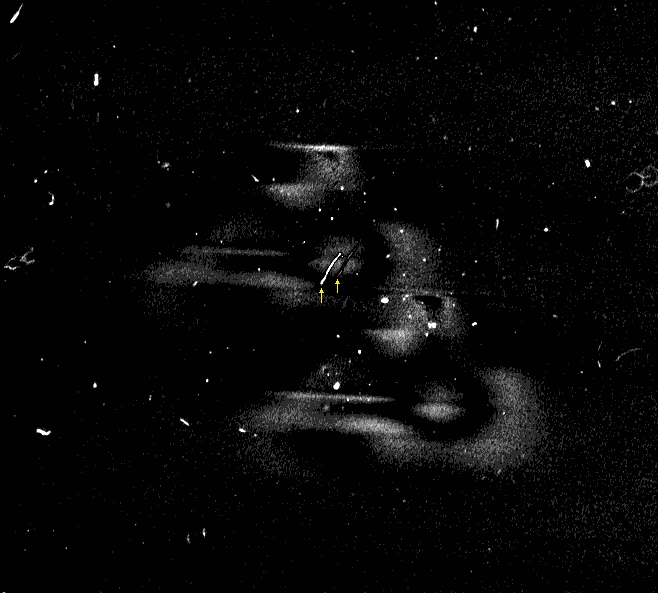}
\caption{Same as in Figure \ref{fa1}, but the brightness in this Figure corresponds to $40 L_{i,j}/\max(L_{i,j})$, capped at 1. Yellow arrows indicate the locations where the shifted dust particle is manifested in the pixelwise loss.}
\label{fa2}	
\end{center}
\end{figure}

\begin{figure}
\begin{center}
\includegraphics[width=120mm]{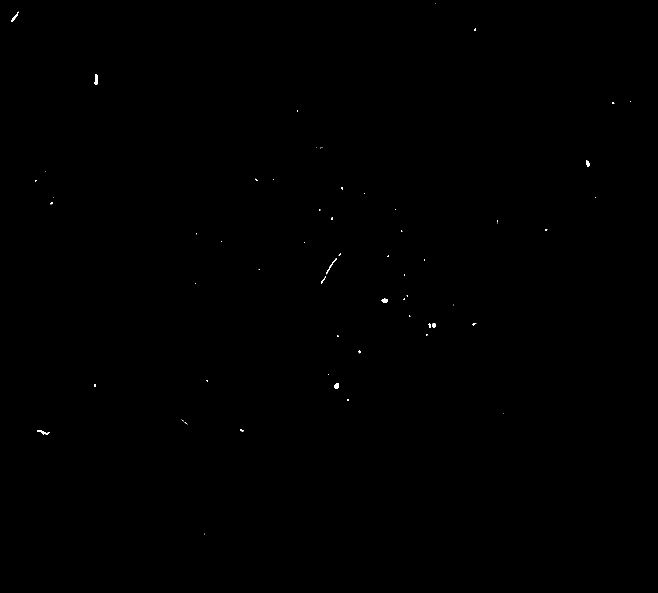}
\caption{Same as in Figure \ref{fa1}, but the brightness in this Figure is 1 is $10 L_{i,j}/\max(L_{i,j}) > 0.25$, and 0 otherwise. Introducing the threshold value allows to definitively locate the impurities on the spectrogram.}
\label{fa3}	
\end{center}
\end{figure}

\begin{figure}
\begin{center}
\includegraphics[width=120mm]{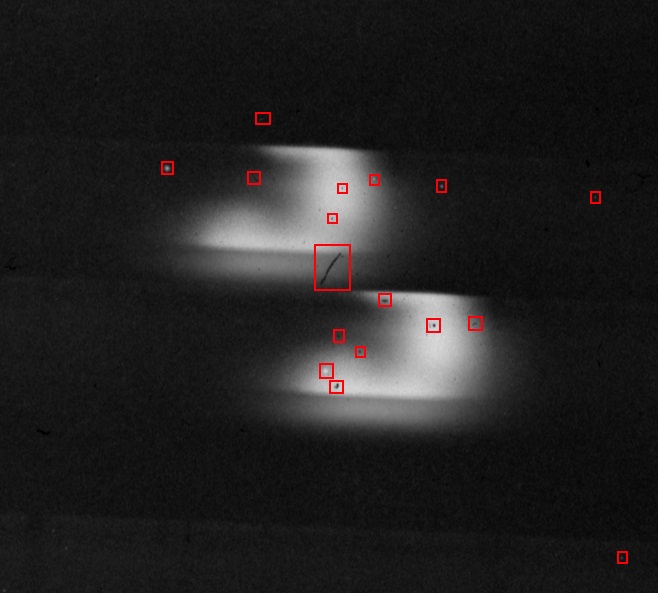}
\caption{Same as in Figure \ref{f5}b. The red rectangles show the location of the visible contamination features, which were selected manually according to the standard spectral processing procedure.}
\label{fa4}	
\end{center}
\end{figure}

\begin{figure}
\begin{center}
\includegraphics[width=120mm]{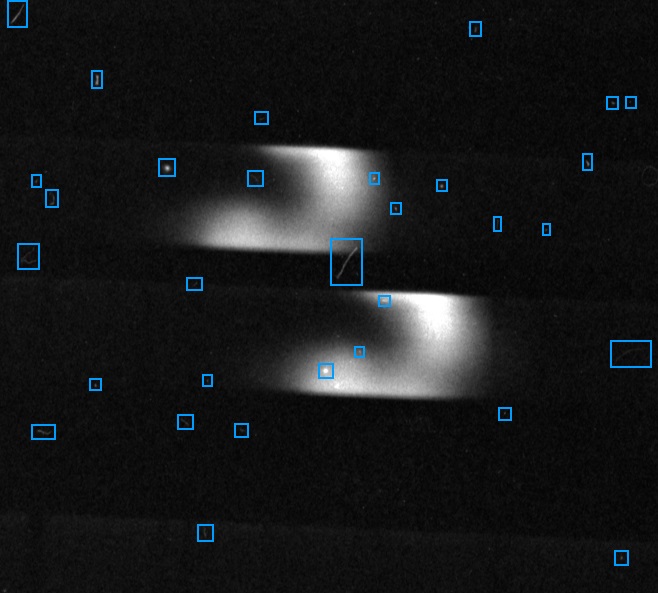}
\caption{Same as in Figure \ref{f5}a. The blue rectangles show the location of the visible contamination features, which were selected manually according to the standard spectral processing procedure.}
\label{fa5}	
\end{center}
\end{figure}

\begin{figure}
\begin{center}
\includegraphics[width=120mm]{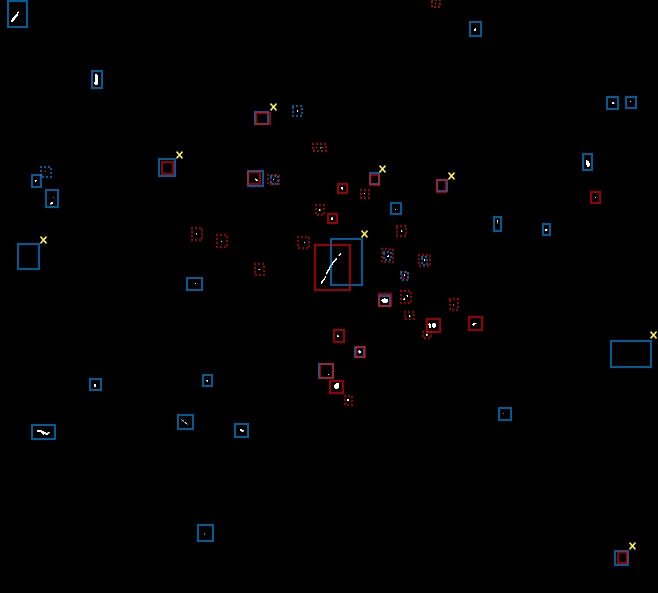}
\caption{Locations of the manually detected features from Figures \ref{fa4} and \ref{fa5} overlaid on Figure \ref{fa3}. Solid rectangles denote the features detected manually. Dotted rectangles show the contaminations that were not initially detected manually, but were verified post-factum. Rectangles marked by yellow crosses correspond to the locations where the manually detected features do not appear in Figure \ref{fa3}; each of these features are present in Figure \ref{fa1}, although with pixelwise loss values not high enough to pass the threshold.}
\label{fa6}	
\end{center}
\end{figure}

\clearpage
\begin{center}
{\bf
Машинне навчання в застосуванні до автоматичної обробки шумів в сонячних спектрограмах}
\end{center}
\begin{center}\bf
І. І. Яковкін$^{a,*}$, О. О. Бартенєв$^{b}$, Н. В. Петрова$^{c}$
\end{center}

\begin{center}
$^{a}$Астрономічна обсерваторія, Київський національний університет імені Тараса Шевченка, Київ 04053, Україна.

$^{b}$Факультет фізики, Університет Пуерто-Рико, Маягуес, Пуерто-Рико 00680, США.

$^{c}$Інститут фізики Національної академії наук України, Київ 03028, Україна.
\end{center}
\begin{center} \bf
Анотація
\end{center}
У статті розроблено підхід з застосуванням машинного навчання для обробки сонячних спектральних даних. Продуктивність запропонованого підходу була продемонстрована на прикладі спектрів лімбового сонячного спалаху 17 липня 1981 року. Результати показали, що таким чином машинне навчання можна ефективно використовувати для виконання перетворення між по-різному оцифрованими спектрами, відновлення унікальних експериментальних даних і проведення очищення спектру. Зокрема, були розроблені згорткові нейронні мережі для перетворення між зображеннями спектрограм сонячного спалаху, отриманими шляхом сканування на просвіт та на відбиття. Запропонований підхід може бути зручним методом обробки спектрограм унікальних сонячних подій, зокрема збільшуючи частку ділянок спектрів, які можна буде використати для подальшого аналізу. Це дозволить у наступних дослідженнях включити до аналізу дані, наприклад такі як ділянки спектрів розташовані близько до країв спектрограм, що раніше не могли бути надійно опрацьованими через технічні обмеження доступних методів обробки. В результаті це призведе до збільшення кількості досліджуваних спектральних ліній для певних спостережуваних подій, що є надзвичайно важливим для побудови фізичних моделей, оскільки спектральні особливості проявляються синхронно в різних спектральних лініях.

Також розроблено протокол для виявлення та усунення недоліків та пошкоджень на спектрограмах. Зазвичай, кожна окрема особливість у спектрі мала бути перевірена вручну, щоб впевнитись у її автентичності, і щоб виключити її з розгляду у випадку якщо це пошкодження, подряпина або частинка пилу. Застосовуючи запропонований протокол для обробки спектрограм, який включає сканування спектрограми за допомогою кількох принципово різних методів, а потім використання машинного навчання для порівняння отриманих зображень, процес виключення домішок можна автоматизувати. Після цього  відбувається процедура відновлення відповідної ділянки спектральної кривої, що дозволяє максимізувати відсоток наявних спостережених даних, що використовуються для подальшого аналізу.

\begin{bf}
Ключові слова: 
\end{bf}
спектри сонячних спалахів, обробка астрономічних зображень, обчислення на графічних процесорах, аналіз астрономічних даних, спектроскопія

\end{document}